\documentclass[aip,twocolumn,reprint]{revtex4-1}
\usepackage{amsmath}
\usepackage{amssymb}
\usepackage{bm}
\usepackage{graphicx}
\usepackage{subcaption}
\usepackage[capitalise]{cleveref}
\usepackage[english]{babel}

\newcommand{\bp}{\ensuremath{\mathbf{p}}}
\newcommand{\bq}{\ensuremath{\mathbf{q}}}
\newcommand{\br}{\ensuremath{\mathbf{r}}}
\newcommand{\bx}{\ensuremath{\mathbf{x}}}

\begin{document}

\title{Do hydrodynamic models misestimate exchange effects? Comparison with kinetic theory for electrostatic waves}
\author{Gert Brodin}
\author{Robin Ekman}
\author{Jens Zamanian}
\affiliation{Department of Physics, Ume{\aa } University, SE--901 87 Ume{\aa}, Sweden}

\date{\today}

\begin{abstract}
We have extended previous quantum kinetic results to compute the exchange correction to the electrostatic electron susceptibility for arbitrary frequencies and wavenumbers in the low temperature limit.
This has allowed us to make a general comparison with a much used hydrodynamic expression, based on density functional theory, for exchange effects.
For low phase velocities, as for ion-acoustic waves, wave-particle interaction leads to a strong enhancement of the exchange correction and the hydrodynamic result is smaller by an order of magnitude.
The hydrodynamic expression gives a useful approximation when the phase velocity is  $\gtrsim 2.5$ times the Fermi velocity.
If this condition is not fulfilled, the hydrodynamical theory gives misleading results.
We discuss the implications of our results for model choice for quantum plasmas, especially regarding particle dispersive effects.
\end{abstract}

\maketitle

\section{Introduction}

A majority of plasmas have low densities and/or high temperatures which assures that they can be accurately described using classical theories.
Recently, however, there has been an increasing interest in plasmas of low temperatures and high densities, which make quantum properties, such as exchange and particle dispersion, significant.
Reviews of recent developments are given in, e.g., Refs.~\onlinecite{Haas-book,manfredi2005model,Shukla-Eliasson-RMP}. 
Much of the research is motivated by applications such as quantum wells \cite{Manfredi-quantum-well}, spintronics~\cite{wolf2001spintronics} and plasmonics~\cite{atwater2007plasmonics}.
Experiments with solid density targets~\cite{glenzer-redmer} are also relevant in this context. 
Dense plasmas are usually divided into strongly and weakly coupled, and into degenerate and non-degenerate electrons~\cite{bonitz-1998}.
Here we will concentrate on the fully degenerate but weakly coupled case.

Exchange and particle dispersive effects both scale as the quantum parameter \begin{equation} 
	H^2 = \frac{\hbar ^2\omega_{pe}^2}{m_e^2v_F^4}
\end{equation}
relative to classical terms.
Here $\hbar $ is the reduced Planck constant, $\omega_{pe}$ is the electron plasma frequency, $m_e$ is the electron mass and $v_F$ is the Fermi velocity.
Many works studying this regime have focused on particle dispersive effects, e.g. using the kinetic Wigner-Moyal equation~\cite{Shukla-Eliasson-RMP,eliassonshukla2010} or by introducing the Bohm-de Broglie term to a fluid model~\cite{Haas-book}.
Since the scaling is the same, including the former effect but not the latter can be questioned.
A possible justification for dropping exchange effects is that the overall $H^2$-dependence comes with a dimensionless factor that depends on the detailed field configuration, and that this factor can be smaller than unity.

From a pragmatic point of view, exchange effects are often neglected because they are complicated to model.
Still there exists a popular quantum hydrodynamic model that includes
not only exchange effects, but also the effect of electron correlations.
It was introduced in Ref.~\onlinecite{crouseilles2008quantum} and has been much used since then \cite{DFT-HF-LF,DFT-LF,DFT-rel,DFT-semicond,haque2018impact,rozina2018raman,sahu2018nonlinear,haque2018dipolar,Chowdhury2017,hussain2018magnetosonic}. 
While this model is straightforward to apply, it should be stressed that the exchange potential is derived using standard density functional theory (DFT), rather than its time dependent counterpart (TDDFT).
Thus it is far from clear that the model has reasonable applicability for dynamic problems, and the omission of kinetic effects means that there are further question marks regarding the accuracy of the model. 

The purpose of the present work is to use a quantum kinetic model derived from first principles~\cite{zamanian2013exchange} to evaluate the accuracy of the above mentioned hydrodynamic model of exchange effects.
Previous attempts in this direction~\cite{ekman2015exchange} have been limited to the long wavelength regime (with the phase velocity much larger than the Fermi velocity) for high wave frequencies, or to the quasi-neutral case for low wave frequencies.
Here we remove these restrictions, and make a general comparison of the quantum kinetic and hydrodynamic models, for linearized electrostatic waves.
Due to the complexity of the kinetic theory, the evaluation must be done under the assumption that the exchange term can be treated as a small perturbation.

The main conclusion is that the hydrodynamic model gives a decent approximation of the exchange effects for phase velocities $\omega/k \gtrsim 2.5 v_F$, but becomes inaccurate for shorter wavelengths.
In particular, for high-frequency waves, the hydrodynamical model predicts the wrong sign of the exchange contribution for short wavelengths.
Moreover, in the low-frequency regime, there is a rather pronounced enhancement of the exchange effects due to wave-particle interaction, which has no counterpart in hydrodynamic theory.
As a result, the hydrodynamic theory underestimates the exchange effects by an order of magnitude, and also the hydrodynamic theory does not capture the variation with phase velocity. 
The enhanced exchange effects in the low-frequency regime has profound implications for the modeling of ion-acoustic waves, which is discussed in some detail in the concluding parts of the paper.       

\section{The kinetic and hydrodynamic dispersion relations}

\subsection{The quantum kinetic case}

Our calculations here are based on the evolution equation for the Wigner-function first derived in Refs.~\onlinecite{zamanian2013exchange,Zamanian2015}.
In that paper the first level in the BBGKY-hierarchy was applied, and the two-particle density matrix was written as an anti-symmetric product of one-particle density matrices.
Neglecting two-particle correlations a correction due to exchange effects was obtained.
In the absence of spin polarization (summing over all spin states) and making long-scale approximations (assuming spatial scale lengths longer than the characteristic de Broglie wavelength), the following evolution equation was derived:
\begin{widetext}
\begin{multline}
 \partial_{t} f(\bx, \bp, t) + \frac{\bp}{m}\cdot \nabla_x f(\bx, \bp, t) - e\mathbf{E}(\bx, t) \cdot \nabla_p f(\bx, \bp, t) = \\
 \frac{1}{2} \partial_p^i \int d^3r d^3q \, e^{-i\br \cdot \bq /\hbar}
  [\partial_{r^{i}}V(\br)]
  f\left(\bx - \frac{\br}{2}, \bp + \frac{\bq}{2}, t\right)
  f\left(\bx - \frac{\br}{2}, \bp - \frac{\bq}{2}, t\right) 
 \\
 -\frac{i\hbar}{8} \partial_p^i \partial_p^j \cdot \int d^3r d^3q \, 
 e^{-i\br \cdot \bq / \hbar}
 	[\partial_r^i V(\br)]\left[ f\left( \bx - \frac{\br}{2}, \bp - \frac{\bq }2,t\right)
 	\left( \overset{\leftarrow}{\partial_x^j} - \overset{\rightarrow}{\partial_x^j} \right)
 	f\left( \bx -\frac{\br}{2}, \bp +\frac{\bq}{2}, t\right)
 	\right].
 	\label{eq:kinetic-eq}
\end{multline}
\end{widetext}
The system is closed with Poisson's equation which takes its classical form, \begin{equation}
	\nabla \cdot \mathbf{E} = -\frac{e}{\epsilon_0} \int d^3\bp \,  f \label{eq:poisson-kin}.
\end{equation}
As is evident, the left hand side in~\cref{eq:kinetic-eq} is the classical Vlasov operator for electrostatic fields, and it should be noted that the expression would be slightly more complicated in the presence of electromagnetic fields~\cite{Zamanian2015}.
The right hand side corresponds to the exchange contribution.
Here we use $\bx$ and $\br$ for position vectors and $\bp $ and $\bq $ for momentum vectors, $e$ is the elementary charge, arrows indicate in which direction an operator acts, and the summation convention is used.
Finally $V(\br) = \frac{e^2}{4\pi \varepsilon_0 \left|\br\right|}$ is the Coulomb potential.

In Ref.~\onlinecite{zamanian2013exchange} \cref{eq:kinetic-eq} was applied to investigate the exchange contribution to linear electrostatic waves in a homogeneous electron-ion plasma with degenerate electrons.
The (hydrogen) ions were assumed to be cold and treated classically.
The exchange effect was treated as a small correction, such that first order perturbation theory was applicable.
Based on \cref{eq:kinetic-eq,eq:poisson-kin} and the conditions given above, Ref.~\onlinecite{ekman2015exchange} derived a dispersion relation 
\newcommand{\electron}{\textnormal{(e)}}
\newcommand{\ion}{\textnormal{(i)}}
\newcommand{\exchange}{\textnormal{(x)}}
\begin{equation}
	1 + \chi^{\ion} + \chi^{\electron}  + \chi^{\exchange} = 0
	\label{eq:disprelation}	
\end{equation}
where the terms are, in order, the cold classical ion susceptibility, the classical electron susceptibility, and the exchange correction to the susceptibility.
They are given by
\begin{align}
	\chi^{\ion} & = -\omega_{pi}^2 / \omega ^2 \label{eq:chii} \\
	\chi^{\electron} & = \frac{3\omega_{pe}^2}{2k^2 v_F^2}\int_{-1}^{1}
	\frac{z \, dz}{z-\omega /kv_F} 
	\label{eq:chie} \\
	\chi^{\exchange} & = (9 \hbar ^2\omega_{pe}^4 / 16m^2k^2v_F^{6}) I(\alpha) 
	\label {eq:chix} \\
	I(\alpha) & = \int_{-1}^{1} dx \int_{-1}^{1} dy \,
	\frac{xy}{\alpha - y} \frac{\operatorname{sgn}(x-y)}{\left( \alpha -(x+y)/2\right)^2}
	\label{eq:integral}
\end{align}
where $\alpha =\omega /kv_F$ is the normalized phase velocity.
When evaluating $I(\alpha)$ for $\omega /kv_F<1$, the Landau prescription for the poles should be used (i.e., we add a small imaginary part $i\delta$ to $\omega$ and take the limit $\delta \to 0$).
When deriving \cref{eq:integral}, no assumption regarding the value of $\omega/kv_F$ was made, i.e., the given dispersion relation applies to both ion-acoustic waves ($\omega /kv_F\ll 1$) and Langmuir waves ($\omega /kv_F > 1$).

Due to the relative complexity of the integral $I(\alpha)$, only two cases were computed in Ref.~\onlinecite{ekman2015exchange}:
the long wavelength regime of Langmuir waves when  $\alpha \gg 1$; and the quasi-neutral limit of ion-acoustic waves when $\alpha = c_{s} / v_F = \sqrt{m_e/3m_{i}}$, where $c_{s} = v_F\sqrt{m_e/3m_{i}}$ is the ion-acoustic velocity in the limit of fully degenerate electrons.
Below we will generalize this treatment and compute the exchange-integral $I(\alpha)$ for the general ranges $0 < \alpha \leq  \sqrt{m_e/3m_i}$ (ion-acoustic waves) and $1<\alpha <\infty $ (Langmuir waves).
Note that for completely degenerate electrons, $\alpha >1$ for Langmuir waves, which means that we will have no Landau damping for this wave mode~\footnote{This is modified by particle dispersive effects, as described by the Wigner equation, and there is Landau damping in the short scale limit~\cite{eliassonshukla2010}.}.

\subsection{The quantum hydrodynamic case}
 Exchange effects can be incorporated in a quantum hydrodynamic framework based on density functional theory~\cite{manfredi2005model,brey1990DFT}.
Using dimensionless variables ($t\to \omega_{pe}t$, $\br\to \omega_{pe}\br/v_F$, $n\to n/ n_{0}$, $\mathbf{v}\to  \mathbf{v}/v_F$, $\phi \to |e|\phi /mv_F^2$), where $n_{0}$ is the unperturbed background density and $\phi $ the electrostatic potential, the model consists of the continuity and momentum equations for electrons,
\begin{equation}
	\partial_t n + \nabla\cdot (n \mathbf{v}_e) = 0 \label{eq:contin-hyd}
\end{equation} \begin{multline}
	\left( \partial_t + \mathbf{v}_e\cdot \nabla \right)  \mathbf{v}_e = \\
	\nabla \big( \phi +
	 \underbrace{%
	 	0.985\frac{\big(3\pi^2\big)^{2/3}\hbar^2\omega _{pe}^2}{4\pi m_e^2v_F^4}n_e^{1/3}
	}_\text{exchange} \big)
	  -\frac{1}{5n}\nabla n^{\gamma } , \label{eq:DFT-eq}
\end{multline}
and Poisson's equation, again taking its classical form, \begin{equation}
	\nabla^2 \phi = n. \label{eq:poisson-hyd}
\end{equation}
When deducing the dimensionless form of the momentum equation we have assumed that the Fermi pressure can be written $P = P_0(n/n_0)^{\gamma}$, with the unperturbed Fermi pressure $P_0 = mv_F^2/5$.
For the case of collision-free electrons, it is natural to determine the exponent $\gamma$ from a comparison with kinetic theory.
We note that we get agreement with the Langmuir dispersion relation in the long wavelength limit if we let $\gamma = 3$, and hence we will use this value in what follows.
The numerical coefficient $0.985$ in the exchange contribution is determined from time-independent DFT~\cite{brey1990DFT}.

When writing down \cref{eq:DFT-eq} it should be noted that we have omitted two terms often kept in quantum hydrodynamic theories.
First, we have dropped the Bohm-de-Broglie term, which is justified if we consider spatial scale lengths longer than the characteristic de Broglie wavelength.
Second, we have omitted a contribution to the effective potential from electron-electron correlations.
While this term scales slightly differently with the electron number density, as compared to exchange effects, in practice these two terms can often be comparable in magnitude~\cite{crouseilles2008quantum}.

The motivation for still dropping correlation effects is two-fold.
Firstly, our aim is to compare with quantum kinetic theories that have only included exchange effects, and thus we do not want to complicate the picture  by keeping physical effects that are irrelevant to the comparison.
Secondly, our basic assumption is that exchange and correlation effects are both small, such that they can be treated perturbatively.
To first order, perturbations can thus be treated independently and
their respective contributions to the dispersion relation can be summed.

\newcommand{\electronhyd}{\textnormal{(e,h)}}
\newcommand{\exchangehyd}{\textnormal{(x,h)}}
\newcommand{\hyd}{\textnormal{h}}

Considering electrostatic wave propagation around a homogeneous equilibrium, linearizing ~\cref{eq:contin-hyd,eq:DFT-eq,eq:poisson-hyd}, and including also the classical ion contribution to the charge density (in the zero temperature limit), we can write the hydrodynamic version of the dispersion relation in the same form as~\cref{eq:integral}.
Naturally the ion susceptibility is the same as in the previous case, \cref{eq:chii}, but now the classical part of the electron susceptibility is  \begin{equation*}
 \chi^{\electronhyd} = -\frac{\omega_{pe}^2}{\omega^2 - (3/5)k^2 v_F^2}
 \end{equation*}
and the electron exchange contribution is  \begin{align}
	\chi^{\exchangehyd} = &\frac{0.985\left( 3\pi^2 \right)^{2/3}\hbar ^2 k^2\omega_{pe}^4}{12\pi m_e^2v_F^2(\omega^2 - (3/5) k^2 v_F^2)^2} \notag \\
	= &\frac{9\hbar^2\omega_{pe}^4}{16 m^2 k^2 v_F^6 }I^{\mathrm{h} }(\alpha) \end{align} 
	where  \begin{equation}
	I^{\mathrm{h}}(\alpha) = \frac{0.445}{(\alpha^2 - (3/5))^2} \label{eq:I-hyd}.
\end{equation}
Here we have denoted the hydrodynamic quantities with an index ``h" to distinguish them from their quantum kinetic counterparts.
When deriving the hydrodynamic expression, the exchange correction appears as a modification of the classical electron susceptibility.
Since the exchange contribution is taken to be small, however, we can Taylor expand the modified susceptibility such that the exchange effect end up as a separate term, in agreement with \cref{eq:disprelation}.

As is well-known, the classical electron susceptibility differs somewhat in the hydrodynamic and kinetic descriptions.
However, we note that with our choice of $\gamma = 3$ in the equation of state both the hydrodynamic and kinetic electron susceptibility can be written as $\simeq - (\omega_{pe}^2 + (3/5)k^2 v_F^2)/\omega^2$ in the long wavelength limit $ k^2v_F^2\ll \omega^2$.
Moreover, when taking out the common factor $9\hbar ^2 \omega_{pe}^4/16m^2k^2v_F^{6}$ from the exchange susceptibility, both the kinetic and hydrodynamic expressions depend on a single function of the $I(\alpha)$ and $I^{\hyd}(\alpha)$ respectively, of the normalized phase velocity $\alpha$.

\subsection{Comparison of $I(\protect\alpha)$ and $I^{\hyd}(\protect\alpha)$}

\begin{figure*}
	\begin{subfigure}{0.47\textwidth}
		\includegraphics[width=\textwidth]{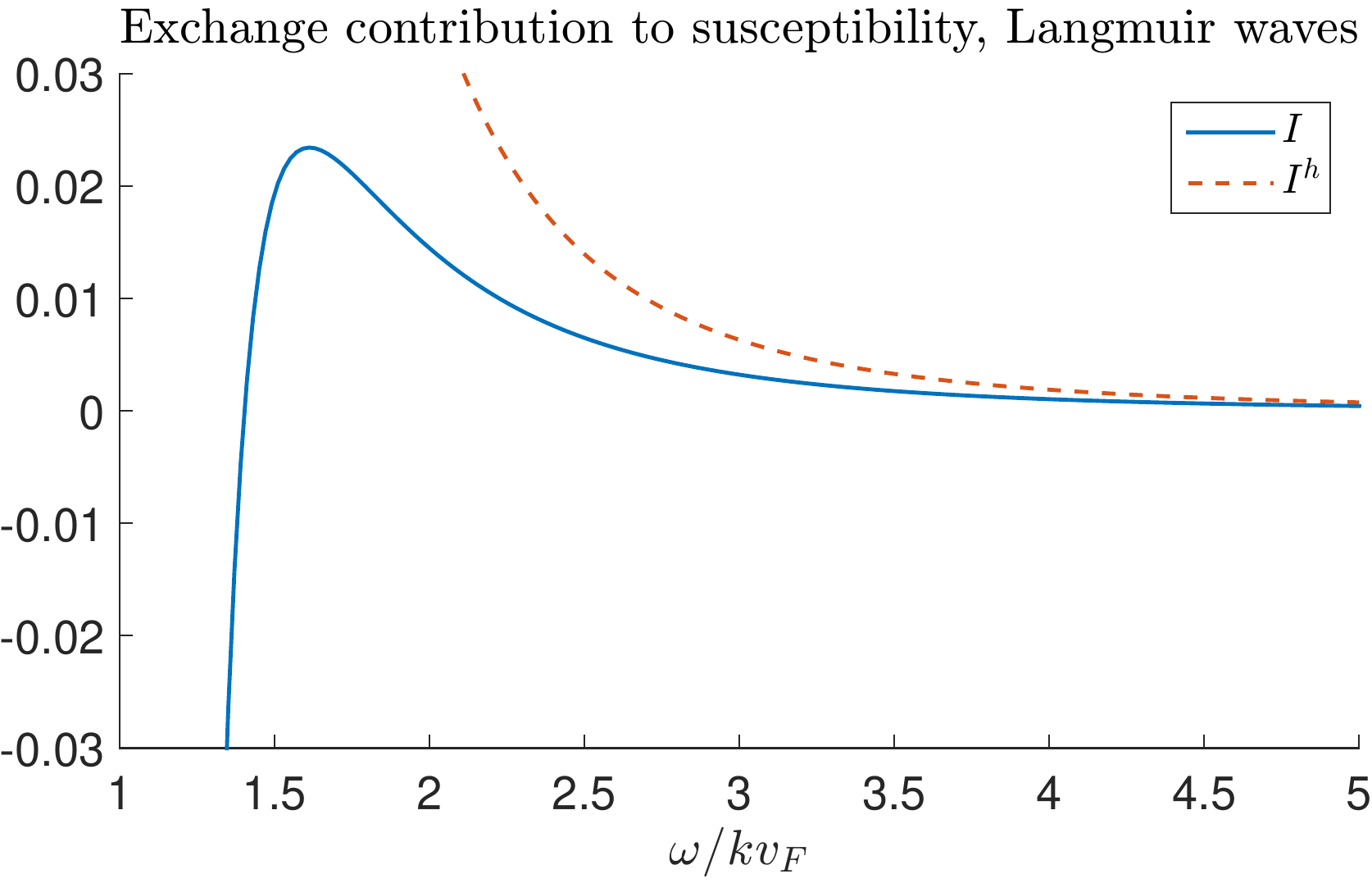}
		\caption{ \label{fig:langmuir-full}}
	\end{subfigure}
	\begin{subfigure}{0.47\textwidth}
		\includegraphics[width=\textwidth]{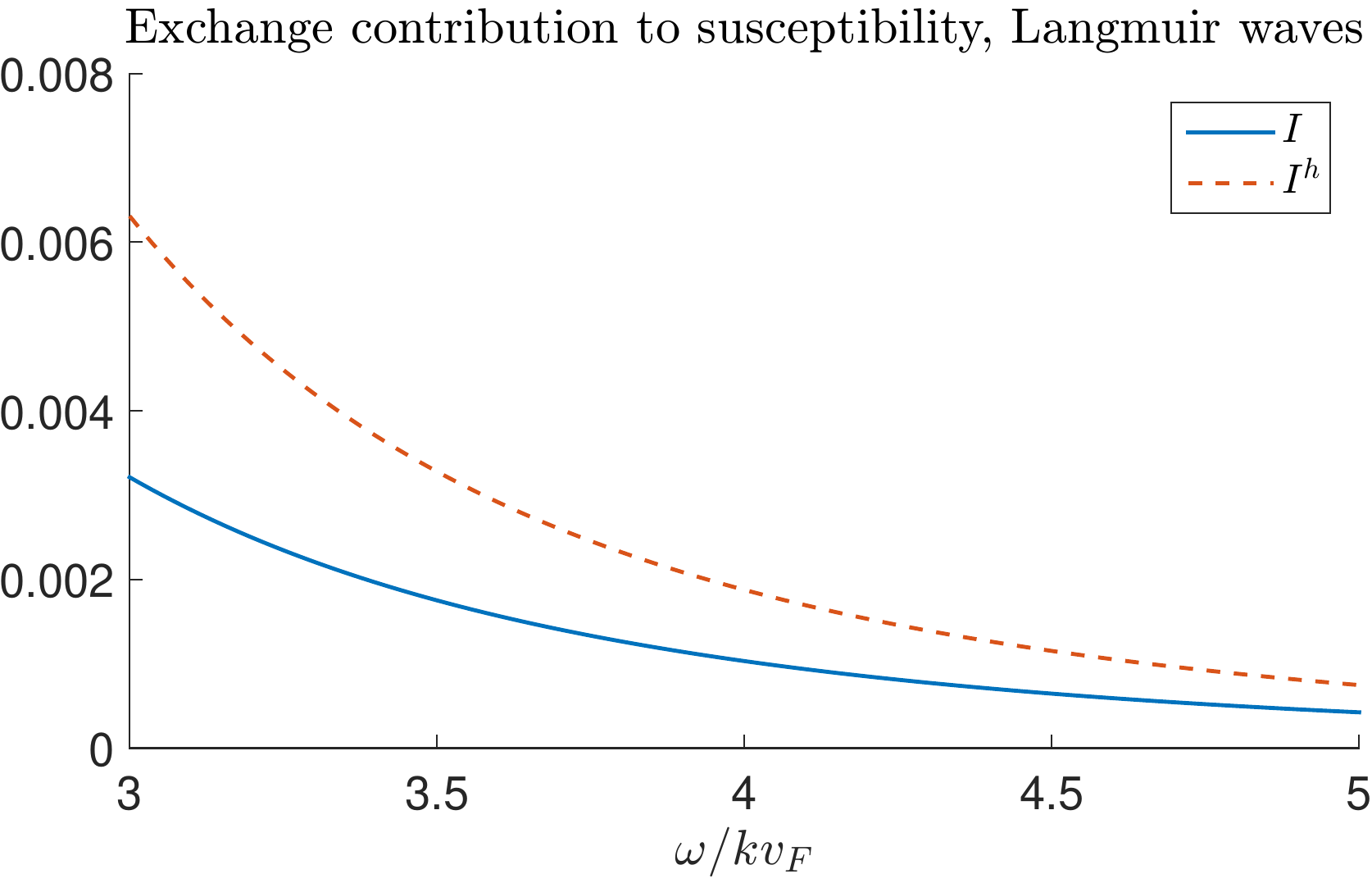}
		\caption{\label{fig:langmuir-zoomed} }		
	\end{subfigure}
	\caption{Exchange contribution to the susceptibility, kinetic (solid) and hydrodynamic (dashed), for Langmuir modes.
	\subref{fig:langmuir-full} Full range of $\alpha$
	\subref{fig:langmuir-zoomed} Detail of the high-frequency regime, showing different limits for $I$ and $I^\hyd$.
\label{fig:langmuir}}
\end{figure*}

\begin{figure}
	\includegraphics[width=\columnwidth]{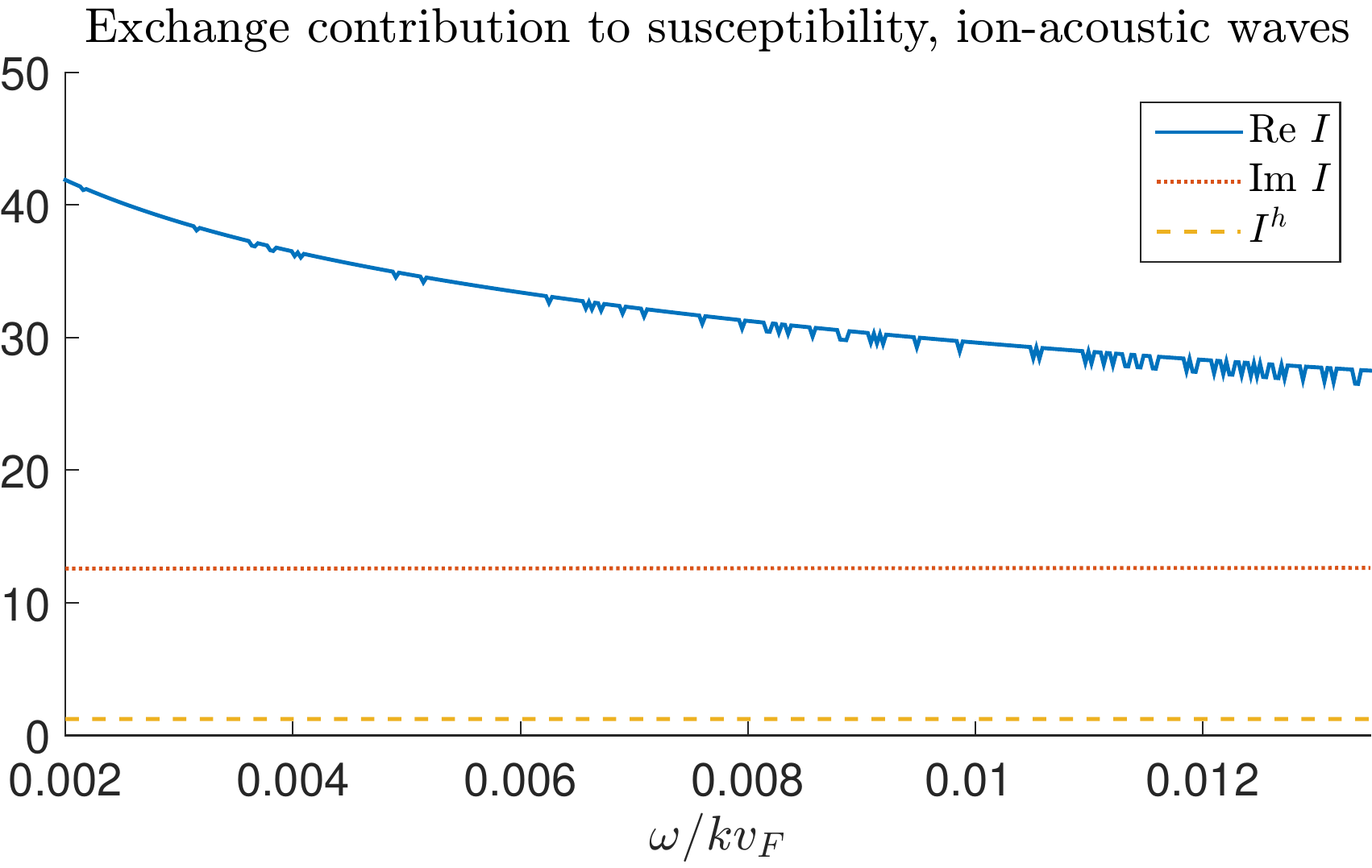}
	\caption{The exchange contribution to the susceptibility, kinetic (solid) and hydrodynamic (dashed), for ion-acoustic modes.
	Note that in the kinetic case, there is also an imaginary part (dotted).
	The irregularities in the real part of $I(\alpha)$ are due to the numerical resolution.
	\label{fig:ionac}}
\end{figure}

From the definitions of the two functions $I(\alpha)$ and $I^{\hyd}(\alpha)$, we see that they coincide if and only if the kinetic and fluid-DFT contributions to the exchange effects are the same.
Thus when comparing $I(\alpha)$ and $I^{\hyd}(\alpha)$, we view the result as an evaluation of the validity of the approximations inherent in the fluid formalism. 
However, the well-known phenomenon of wave-particle interaction (leading to imaginary pole contributions in $I(\alpha)$ when $\alpha < 1 $) means that a perfect agreement is ruled out from the start.
Nevertheless it is possible that the fluid DFT expression for the exchange effects can serve as a useful approximation under suitable conditions.
For an electron-proton plasma we do not need to investigate the full range of $\alpha$.
Either $1 < \alpha < \infty $ (the Langmuir mode) or $0< \alpha < \sqrt{m_e/3m_i} \approx 0.013 $ (the ion acoustic mode).
While in principle we may also have a damped electron-acoustic mode~\cite{valentini2012undamped} with $\alpha \lesssim 1$, it is not meaningful to compare $I^{\hyd}(\alpha)$ and $I(\alpha)$ in this case, as the electron-acoustic mode depends critically on a kinetic description and does not even exist in fluid theory.

In \cref{fig:langmuir-full}, we have evaluated the integral in \cref{eq:integral} numerically, and plotted $I(\alpha)$ and $I^{\hyd}(\alpha)$ for $1.1 < \alpha <5$.
Two interesting effects that have not previously been reported should be noted.
First of all, for $\alpha \approx 1.4$, the exchange contribution to the susceptibility becomes negative.
Secondly,  when $\alpha \to 1$ the fluid resonance tends to increase $I^{\hyd}(\alpha)$.
This is in stark contrast to the behavior seen in fluid theory, where not even the sign agrees for small $\alpha$. 

The large discrepancy for small $\alpha $ is related to the different role of resonances in fluid and kinetic theory.
When $\alpha \to 1$ the fluid resonance tends to increase $I^{\hyd}(\alpha)$, but the kinetic effects act in the opposite direction and even lead to a sign change of $I(\alpha)$.
Formally, the integral diverges when $\alpha \to 1$, however, when $I(\alpha) H^2 \sim 1$ our perturbative treatment breaks down.
One needs to stress here that the kinetic theory avoids many of the approximations inherent in the fluid treatment.

From \cref{fig:langmuir-full} it may seem that the agreement is very good for the larger values of $\alpha$.
However, zooming in on the long-wavelength regime as in~\cref{fig:langmuir-zoomed}, we see that there is still a significant discrepancy.
We can understand this by considering the asymptotics.
In the limit of $\alpha \gg 1$ the integral $I(\alpha)$ can be evaluated analytically and one finds $I(\alpha) \approx \frac{4}{15}\alpha^{-4}, \alpha \gg 1$; this agrees exactly with previous works using several different methods~\cite{nozierespines1958,vonRoos1961,kanazawa1960}.
The hydrodynamic expression~\cref{eq:I-hyd} has the same $\alpha^{-4}$ scaling for large $\alpha$, but with a numerical coefficient that is approximately a factor $5/3$ larger, and this is what is seen in~\cref{fig:langmuir-zoomed}.

Next we turn to ion-acoustic waves, where $\omega /k = c_{s}$ in the long wavelength limit when the quasi-neutral limit applies, but for shorter wavelengths $\omega/k < c_s$.
In any case $\alpha <1$, and we will therefore have pole contributions to $I(\alpha)$, giving a complex susceptibility.
The imaginary part can be viewed as an exchange modification of the Landau damping rate.
Naturally this has no correspondence in the fluid description, but we can still compare the real parts.

In \cref{fig:ionac} we show the real and imaginary parts of $I(\alpha)$, and $I^{\mathrm{h}}(\alpha)$ in the ion-acoustic range where $0<\alpha < 0.013$.
While the real parts of $I(\alpha)$ and $I^{\hyd}(\alpha)$ are both positive, we see that we are now very far from agreement.
Importantly $I(\alpha)$ is an order of magnitude larger, and shows a significant dependence on $\alpha$, whereas $I^{\hyd}(\alpha)$ is approximately constant in the ion-acoustic regime.

We note that the integral $I(\alpha)$ has large contributions from the regions where the denominators are small, corresponding to exchange-mediated wave-particle interaction.
Since wave-particle interaction is important whenever $\alpha < 1$, it is not entirely surprising that the fluid model becomes inaccurate in this regime.
This is further emphasized by the rather large imaginary part of $I(\alpha)$.
The imaginary part shows very little variation with $\alpha$, but the large magnitude means that the exchange-modification of the Landau damping rate can be significant even for a modest value of the quantum parameter $H^2$ that determines the  importance of exchange effects, see~\cref{eq:chix}.

Since wave-particle interaction does not take place for the Langmuir mode, the large deviation of the fluid and kinetic results for short wavelengths may be somewhat surprising. 
In particular for $\alpha\lesssim1.5$, the hydrodynamic theory is seriously misleading. 
The large discrepancy is related to the different role of resonances in fluid and kinetic theory.

\section{Discussion}

In the present paper we have relaxed the assumptions made in a previous work \cite{ekman2015exchange} to give a kinetic treatment of exchange effects covering the full spectrum of electrostatic waves. 
The new effects reported include: 
Firstly, a change of sign of the exchange susceptibility for Langmuir waves when $\alpha\approx1.4$. 
Secondly, a strong enhancement of the magnitude of the exchange contribution for Langmuir waves in the short wavelength limit.
Thirdly, an increase of the exchange susceptibility of ion-acoustic waves for short wavelengths.  

An important goal of the present investigation has been to examine the validity of a much used (see, e.g., Refs.~\onlinecite{DFT-HF-LF,DFT-LF,DFT-semicond,haque2018impact,rozina2018raman,sahu2018nonlinear,haque2018dipolar,Chowdhury2017,hussain2018magnetosonic}) quantum hydrodynamic expression for exchange effects.
The conclusion is that for linear theory, the hydrodynamic expression provides a useful (but far from precise, see~\cref{fig:langmuir-zoomed}) approximation for high-frequency waves with phase velocities $\omega/kv_F \gtrsim 2.5$.
While the investigation is based on linear theory alone, presumably this may generalize also to nonlinear scenarios.
However, for shorter wavelengths, the hydrodynamic and kinetic treatments disagree dramatically, and even predicts opposite signs of the exchange for the case of Langmuir waves when $\alpha<1.4$.

For low-frequency dynamics with small phase velocities $\omega /kv_F \ll 1$, wave-particle interaction is important, and the hydrodynamic theory underestimates the importance of exchange effects by at least an order of magnitude.
This failure is in addition to the obvious neglect of wave-particle damping mechanisms.
While it is well-known that wave-particle interaction is potentially important for low-frequency phenomena, the difficulty of modeling exchange effects in a quantum kinetic framework has led many researchers to adopt the fluid model presented here also for the case of low-frequency exchange dynamics, see. e.g. Refs.~\onlinecite{DFT-HF-LF,DFT-LF,haque2018impact,rozina2018raman,sahu2018nonlinear,haque2018dipolar,Chowdhury2017,hussain2018magnetosonic}.
Our comparison with kinetic theory, however, shows that such an approximation should be avoided.

The unexpectedly large magnitude of exchange effects for low-frequency phenomena has a broader significance.
Many theoretical papers study quantum effects in plasmas concentrating on the effect of particle dispersion.
This can be done quantum kinetically through the use of the Wigner equation, or quantum hydrodynamically by including the Bohm-de Broglie term.
In any case the quantum effects of this type are important for short scale lengths, with wavenumbers of the order $\hbar k \sim mv_F $.
However, if the particle density is low (with $\omega_{pe}^2 \ll k^2 v_F^2$) collective plasma behavior will be more or less negligible, and essentially the particles will be free-streaming (or possibly governed by an external field) Thus for plasma effects to be important we also need a sufficient density $\omega_{pe}^2 \sim k^2 v_F^2$.
For scales short enough to make the particle dispersive effects important (i.e., $k\sim mv_F/\hbar $) we therefore must have $\hbar^2\omega_{pe}^2 / m_e^2 v_F^4 \sim 1$.
With $H^2 = \hbar^2\omega_{pe}^2/m^2v_F^4\sim 1$, the exchange contribution to the susceptibility is still rather small for $\alpha >1$, except very close to unity. 
Although the exchange susceptibility tends to increase somewhat when $\alpha \to 1$, it is still a decent approximation to ignore exchange effects even for $H^2\sim 1$.
Hence it makes sense to describe high-frequency Langmuir waves including particle dispersive effects but dropping exchange effects.
Naturally, high accuracy still requires the inclusion of the exchange contribution, but if only modest precision is required, such an approximation can be justified.

However, for low-frequency ion-acoustic phenomena, the situation is different due to the large value of $I(\alpha)$, see \cref{fig:ionac}.
Since the exchange effects scales as $I(\alpha)H^2$, see \cref{eq:chie,eq:chix}, for low-frequency phenomena these quantum effects dominates over particle dispersive effects, whose contribution is limited by $H^2$.
Thus for ion-acoustic waves it does not make sense to use the Wigner equation for electrons, or quantum hydrodynamical equations with the Bohm term, without an exchange contribution.

In the above context it is a fundamental problem that the quantum kinetic theory is so complicated that it can be only solved perturbatively, and for relatively simple cases (like linear theory for homogeneous media).
By contrast the fluid theory is fairly straightforward to use even for complicated nonlinear problems, but as shown here it underestimates the importance of exchange effects severely.
A conceivable solution when describing low-frequency phenomena would be to use the DFT-expression but with a numerical coefficient increased according to $0.985 \mapsto 20$, compare \cref{fig:ionac} 
\footnote{In Ref.~\onlinecite{ekman2015exchange} a substitution with a somewhat smaller factor was mentioned in a similar context.
It should be noted that the figure is depending on the polytropic factor $\gamma $ in the equation of state: $(P/P_{0}) = (n/n_{0})^{\gamma }$.
If we replace our value $\gamma = 3$ with $\gamma = 5/3$ which is sometimes used~\cite[Section 5.2]{manfredi2005model}, the discrepancy between the fluid model and the kinetic model shown in \cref{fig:ionac} would be reduced by a factor $9/5$ -- i.e., the mismatch would roughly be a factor $10$ instead of a factor $20.$}.

Still, this is not satisfactory, as there is no guarantee such an approach gives a decent approximation beyond this specific case (perturbative treatment of linear ion-acoustic waves). 
Moreover, neither the variations in $I(\alpha)$ will be captured nor the rather large contribution from the imaginary part of $I(\alpha)$.
The general conclusion from this discussion is that it is a major challenge to make an accurate quantitative description of quantum ion-acoustic waves, even for the simple case of linear homogeneous wave propagation.
The descriptions most commonly used (Wigner equation and various quantum hydrodynamical models) either ignore or underestimate the large contribution from exchange effects, and the quantum kinetic theory we have based this work on can only be applied in a perturbative setting.
Thus for the case when $H^2 \sim 1$, all treatments of ion-acoustic waves have serious shortcomings.
A possible way forward could be based on time-dependent density functional theory but finding accurate functionals in the presence of strong wave-particle interaction is likely not straightforward.

\section{Concluding remarks}

Some of the above arguments suggest that the hydrodynamical model based on standard DFT should be avoided for all time-dependent phenomena.
However, that conclusion is potentially too strong.
For high-frequency phenomena in the long wavelength limit, it may still give useful results.
Intuitively one may think that time independent DFT should work better for slowly varying phenomena, but the lack of wave particle interaction (as is usually the case for high frequencies) tends to make the hydrodynamic theories more adequate for rapid processes.
In this context, it should be pointed out that the quantum kinetic models we have used here have several practical limitations.
Most importantly these theories are very hard to apply for nonlinear and/or inhomogeneous problems.
With that said, the general conclusion is that the hydrodynamical DFT based model for exchange effects must be applied with a lot of caution.
Still much more work needs to be done -- experimental and/or theoretical -- in order to get a clear picture of the applicability of the current DFT theories.
On the theoretical side, there is a need for the quantum kinetic formalism presented here to be complemented by works using, e.g., time-dependent density functional theory.

\bibliography{Exchange-short}

\end{document}